\documentclass[aps,amsmath,amssymb, prc,twocolumn,superscriptaddress]{revtex4-1}
\usepackage{graphicx}% Include figure files
\usepackage{dcolumn}% Align table columns on decimal point
\usepackage{bm}% bold math
\usepackage[utf8]{inputenc}
\usepackage[english]{babel}
\usepackage{float}
\usepackage{multirow}
\usepackage{longtable}
\usepackage{dcolumn}
\newcolumntype{d}{D{.}{.}{-1}}
\usepackage{setspace}
\usepackage{threeparttable}
\usepackage{tabularx}
\usepackage[symbol*]{footmisc}
\DefineFNsymbolsTM{myfnsymbols}{% def. from footmisc.sty "bringhurst" symbols
  \textasteriskcentered *
  \textdagger    \dagger
  \textdaggerdbl \ddagger
  \textsection   \mathsection
  \textbardbl    \|%
  \textparagraph \mathparagraph
}%
\setfnsymbol{myfnsymbols}

\usepackage{siunitx}
\newcommand{\nuc}[2]{\hbox{$^{#1}$#2}}

\usepackage{gensymb}
\usepackage{hyperref}

\begin{document}

%\preprint{}

\title{Proton removal from \nuc{73,75}{Br} to \nuc{72,74}{Se} at intermediate energies}


\author{M. Spieker}
\email[]{Corresponding author: mspieker@fsu.edu}
\affiliation{Department of Physics, Florida State University, Tallahassee, Florida 32306, USA}

\author{D. Bazin}
\affiliation{Facility for Rare Isotope Beams, Michigan State University, East Lansing, Michigan 48824, USA}
\affiliation{Department of Physics and Astronomy, Michigan State University, East Lansing, Michigan 48824, USA}

\author{S. Biswas}
\affiliation{Facility for Rare Isotope Beams, Michigan State University, East Lansing, Michigan 48824, USA}

%\author{B.A. Brown}
%\affiliation{Facility for Rare Isotope Beams, Michigan State University, East Lansing, Michigan 48824, USA}
%\affiliation{Department of Physics and Astronomy, Michigan State University, East Lansing, Michigan 48824, USA}

\author{P.D. Cottle}
\affiliation{Department of Physics, Florida State University, Tallahassee, Florida 32306, USA}

\author{P.J. Farris}
\affiliation{Facility for Rare Isotope Beams, Michigan State University, East Lansing, Michigan 48824, USA}
\affiliation{Department of Physics and Astronomy, Michigan State University, East Lansing, Michigan 48824, USA}

\author{A. Gade}
\affiliation{Facility for Rare Isotope Beams, Michigan State University, East Lansing, Michigan 48824, USA}
\affiliation{Department of Physics and Astronomy, Michigan State University, East Lansing, Michigan 48824, USA}

\author{T. Ginter}
\affiliation{Facility for Rare Isotope Beams, Michigan State University, East Lansing, Michigan 48824, USA}

\author{S. Giraud}
\affiliation{Facility for Rare Isotope Beams, Michigan State University, East Lansing, Michigan 48824, USA}

\author{K.W. Kemper}
\affiliation{Department of Physics, Florida State University, Tallahassee, Florida 32306, USA}

\author{J. Li}
\affiliation{Facility for Rare Isotope Beams, Michigan State University, East Lansing, Michigan 48824, USA}

\author{S. Noji}
\affiliation{Facility for Rare Isotope Beams, Michigan State University, East Lansing, Michigan 48824, USA}

\author{J. Pereira}
\affiliation{Facility for Rare Isotope Beams, Michigan State University, East Lansing, Michigan 48824, USA}

\author{L.A. Riley}
\affiliation{Department of Physics and Astronomy, Ursinus College, Collegeville, PA 19426, USA}

\author{M.K. Smith}
\affiliation{Facility for Rare Isotope Beams, Michigan State University, East Lansing, Michigan 48824, USA}

%\author{J.A. Tostevin}
%\affiliation{Department of Physics, Faculty of Engineering and Physical Sciences, University of Surrey, Guildford, Surrey GU2 7XH, United Kingdom}

%\author{A. Volya}
%\affiliation{Department of Physics, Florida State University, Tallahassee, Florida 32306, USA}

\author{D. Weisshaar}
\affiliation{Facility for Rare Isotope Beams, Michigan State University, East Lansing, Michigan 48824, USA}

\author{R.G.T. Zegers}
\affiliation{Facility for Rare Isotope Beams, Michigan State University, East Lansing, Michigan 48824, USA}
\affiliation{Department of Physics and Astronomy, Michigan State University, East Lansing, Michigan 48824, USA}


\date{\today}

\begin{abstract}

We report new experimental data for excited states of \nuc{72,74}{Se} obtained from proton removal from \nuc{73,75}{Br} secondary beams on a proton target. The experiments were performed with the Ursinus-NSCL Liquid Hydrogen Target and the combined GRETINA+S800 setup at the Coupled Cyclotron Facility of the National Superconducting Cyclotron Laboratory at Michigan State University. Within uncertainties, the inclusive cross sections for proton removal from \nuc{73,75}{Br} on a proton target are identical suggesting that the same single-particle orbitals contribute to the proton-removal reaction. In addition, details of the partial cross section fragmentation are discussed. The data might suggest that $l = 1, 2, 3$, and 4 angular momentum transfers are important to understand the population of excited states of \nuc{72,74}{Se} in proton removal. Available data for excited states of \nuc{74}{Ge} populated through the \nuc{75}{As}$(d,\nuc{3}{He})$\nuc{74}{Ge} proton-removal reaction in normal kinematics suggest indeed that the $fp$ and $sd$ shell as well as the $1g_{9/2}$ orbital contribute. A comparison to data available for odd-$A$ nuclei supports that the bulk of the spectroscopic strengths could be found at lower energies in the even-even Se isotopes than in, for instance, the even-even Ge isotopes. In addition, the population of high-$J$ states seems to indicate that multi-step processes contribute to proton-removal reactions at intermediate energies in these collective nuclei.

%The data are compared to existing proton knockout data for \nuc{70}{Se}, which were obtained on a \nuc{9}{Be} target. In the light of recent predictions for nucleon removal reactions on nuclear and nucleon targets for the much lighter nucleus \nuc{29}{Ne}, 

\end{abstract}


\maketitle

%\section*{}
%\subsection{}
%\subsubsection{}

\section{Introduction}

The Ge-Sr isotopes with neutron number $N \leq 40$ are known to feature rapid shape changes with both nucleon number and angular momentum, while also displaying shape coexistence and quadrupole collective, rotational bands \cite{Lec78a, Lec80b, Ebe88a, Kot90a, Tab90a, Cot90a, Hey11a, Gar21a, Gad05a, Cle07a, Lju08a, McC11a, Iwa14a, Goe16a, Wim18a, Hen18a, Wim20a, Wim21a, Muk22a, Bha22a, Ele23a}. Similar observations and predictions have been made for the neutron-rich isotopes of these isotopic chains (see, {\it e.g.}, Refs. \cite{Alb13a, Che17a, Lit18a, Ger22a, Nom22a}). In addition, contributions from both the octupole as well as the hexadecapole degrees of freedom were recently highlighted on the neutron-deficient side \cite{Spi22a,Spi23a}. These higher-order degrees of freedom provide important complementary information on the shape and shell structure of the nuclei in this mass region. While the sudden increase of octupole collectivity in \nuc{72}{Se} and \nuc{70,72}{Ge} might be linked to the prolate-oblate shape transition at $A=72$ and could support the crossing of different microscopic configurations \cite{Spi22a}, the enhanced electric hexadecapole transition strength in \nuc{74,76}{Kr} appears to be connected to the well deformed prolate configuration which dominates the yrast structure at higher angular momenta \cite{Spi23a}. The exact location of the prolate-oblate shape transition is, however, still under debate and its details keep challenging state-of-the-art theoretical models since triaxial degrees of freedom are expected to contribute as well \cite{Lue85a, Fis05a, Guo07a, Gir09a, Rod14a, Nik14a, Aya16a, Hen19a, Aya19a, Aya23a, Mar23a}. Available experimental data suggest that the prolate-oblate ground-state shape transition for even-$A$ nuclei in the Ge-Kr region occurs around neutron number $N = 36$. Many models (see, {\it e.g.}, Refs. \cite{Ben06a, Gir09a, Hin10a, Sat11a, Rod14a}) predict predominantly oblate-deformed ground states for nuclei around $N = 36$, with the yrast structure changing from oblate to prolate with increasing angular momentum. A possible impact of isospin-symmetry breaking effects on the structure and shapes of nuclei close to the shape-transitional point has also been controversially discussed recently \cite{Hof20a, Hen20a, Wim21a, Len21a, Wan22a, Vit22a}.

There is some evidence that the number of protons and which orbits they occupy plays an important role in determining structure changes around the shape-transitional point and that considering the neutron number alone would be too simplistic (see, {\it e.g.}, the discussion around Fig.\,38 in the review article \cite{Hey11a}). Understanding occupancies of proton orbits and the proton-hole structure of excited states could, thus, be essential. In recent decades, nucleon-removal reactions on rare-isotope beams have provided invaluable insights into the single-particle (single-hole) structure of nuclei and the evolution of shell structure on both the neutron-deficient as well as neutron-rich side off the valley of $\beta$ stability \cite{Gad08a, Aum21a, Tos21a}. Systematic studies of inclusive and partial nucleon-removal cross sections for neutron-deficient nuclei in the Ge-Sr region have, however, not been the focus of extensive research so far.

This article focuses on the neutron-deficient, even-even Se isotopes with $A = 70-74$, which were studied through proton-removal reactions from \nuc{71-75}{Br} secondary beams in inverse kinematics. These proton-removal reactions test the occupancies of the different proton single-particle orbits in the respective Br isotopes and the structure overlap with excited states of the Se isotopes. The ground-state spins of the Br isotopes, i.e. of the incoming secondary beams, change in all three cases: $J^{\pi} = 3/2^-$ for \nuc{75}{Br}\,\cite{Eks80a}, $1/2^-$ for \nuc{73}{Br}\,\cite{Gri92a,Mie99a}, and $(5/2^-)$ for \nuc{71}{Br}\,\cite{Wan22b}. In a Nilsson model picture, this sequence could only be explained on the oblate side of quadrupole deformation \cite{Boh98b} if no shape change was to be envoked with changing neutron number. All ground-state configurations would then be dominated by Nilsson configurations originating from the spherical $1f_{5/2}$ proton orbital. In this context, we note, however, that the possibility of observing the prolate-oblate shape transition between the ground states of \nuc{70}{Se} and \nuc{72}{Se} was discussed in, {\it e.g.}, Refs. \cite{Lju08a, Hen19a, Wim18a, Wim21a}. The importance of triaxial degrees of freedom for understanding the structure of \nuc{74}{Se} was emphasized in Ref.\,\cite{Mar23a}. Proton-removal data from \nuc{71}{Br} to \nuc{70}{Se} ($N=36$), obtained through \nuc{9}{Be}(\nuc{71}{Br},X)\nuc{70}{Se} proton knockout at RIKEN, have already been published \cite{Wim18a}. In their work, Wimmer {\it et al.} provided evidence for shape coexistence in \nuc{70}{Se}\,\cite{Wim18a}. In this article, we add the data for \nuc{72,74}{Se} ($N=38,40$) obtained from proton-removal reactions on a proton target measured at the Coupled Cyclotron Facility of the National Superconducting Cyclotron Laboratory at Michigan State University \cite{NSCL}. In addition to the inclusive cross sections, we will discuss the fragmentation of the partial cross sections including the previous data for \nuc{70}{Se}. Using a comparison to $(d,\nuc{3}{He})$ data for stable nuclei in this mass region, we will provide evidence that orbitals from the $fp$ and $sd$ shell, as well as the $1g_{9/2}$ orbital contribute to the population of excited states in \nuc{70-74}{Se}. The bulk of this strength seems to be found at considerably lower energies in the Se isotopes than in, for instance, the Ge isotopes. We will also discuss the possible contribution of multi-step processes to the population of high-$J$ states in proton-removal reactions at intermediate energies in these collective nuclei.

\section{Experiment}

The experiments were performed at the Coupled Cyclotron Facility of the National Superconducting Cyclotron Laboratory (NSCL) at Michigan State University\,\cite{NSCL}. In this work, we present results for the \nuc{73}{Br}$(p,2p)$\nuc{72}{Se} and \nuc{75}{Br}$(p,2p)$\nuc{74}{Se} reactions in inverse kinematics. Results using other components of the secondary-beam cocktail and concentrating on inelastic proton scattering rather than proton removal were already reported in Refs.\,\cite{Spi22a, Spi23a}. The secondary \nuc{73}{Br} (36\,$\%$ purity; $\sim 2160$ pps) and \nuc{75}{Br} (5\,$\%$ purity; $\sim 300$ pps) beams were also produced from a 150\,MeV/u $^{78}$Kr primary beam in projectile fragmentation on a 308-mg/cm$^2$ thick $^{9}$Be target. The A1900 fragment separator\,\cite{Mor03}, using a 240-mg/cm$^2$ Al degrader, was optimized to select \nuc{74,76}{Kr} in flight using two separate magnetic settings. For the magnetic setting centered on \nuc{74}{Kr} and \nuc{76}{Kr}, respectively, the secondary beams \nuc{73}{Br} and \nuc{75}{Br} were part of the cocktail beam. Each secondary beam could be unambiguously distinguished from the other components in the cocktail beam via the time-of-flight difference measured between two plastic scintillators located at the exit of the A1900 and the object position of the S800 analysis beam line. Downstream, the NSCL/Ursinus Liquid Hydrogen (LH$_2$) Target was located at the target position of the S800 spectrograph. The projectilelike reaction residues entering the S800 focal plane, including \nuc{72,74}{Se}, were identified event-by-event from their energy loss and time of flight\,\cite{Baz03}.

\begin{figure}[t]
\centering
\includegraphics[width=1\linewidth]{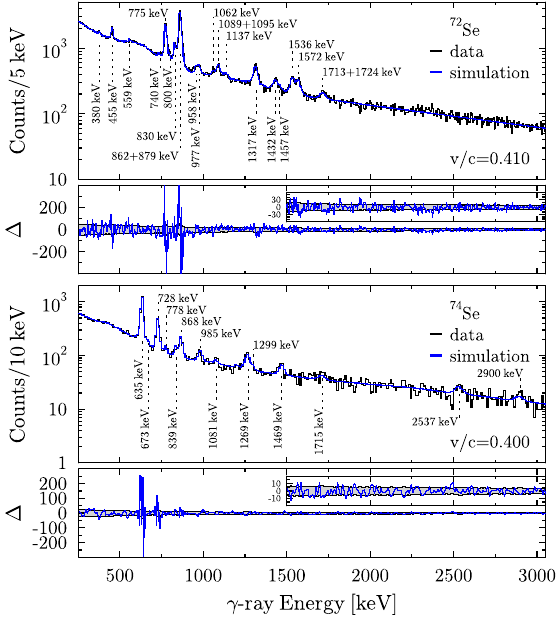}

\caption{\label{fig:spectra}{Doppler-corrected, in-beam $\gamma$-ray spectra for \nuc{72}{Se} (top) and \nuc{74}{Se} (bottom). Data are shown in black. \textsc{geant4} simulations performed with \textsc{ucgretina} \cite{Ril21a} are presented in blue. A prompt background consisting of two exponential functions was included when fitting the simulated spectra to the measured ones. Energies of some resolved $\gamma$-ray transitions used in the simulation are highlighted. More information is provided in Tables\,\ref{table_01} and \ref{table_02}. The weak 2537-keV and 2900-keV transitions could not be placed in the level scheme of \nuc{74}{Se}. The residuals between the measured and simulated spectra are shown below the spectra for \nuc{72}{Se} and \nuc{74}{Se}, respectively. Insets in these show the region between 1500 and 3050\,keV.}} 
 \end{figure}

The GRETINA $\gamma$-ray tracking array \cite{Pas13a,Wei17a} was used to detect $\gamma$ rays emitted by the reaction residues in flight ($v/c \approx 0.4$). Eight GRETINA modules, containing four, 36-fold segmented HPGe detectors each, were mounted in the north half of the mounting shell to accommodate the LH$_2$ target. In this configuration, two modules are centered at 58$^{\circ}$, four at 90$^{\circ}$, and two at 122$^{\circ}$ with respect to the beam axis. At beam velocities of $v/c \approx 0.4$, event-by-event Doppler reconstruction of the residues' $\gamma$-ray energies is key. This reconstruction was performed based on the angle of the $\gamma$-ray emission determined from the main-interaction point and including trajectory reconstruction of the residues through the S800 spectrograph \cite{Wei17a}. Doppler-corrected in-flight $\gamma$-ray spectra are presented in Fig.\,\ref{fig:spectra}. The $\gamma$-ray yields were obtained by fitting $\gamma$-ray spectra, simulated with \textsc{ucgretina} \cite{Ril21a}, to the experimentally observed ones. For these fits, the \textsc{root} \cite{root} \textsc{minuit2} minimizer with the default minimization algorithm \textsc{migrad} was used \cite{minuit2}; as done in previous studies, see, {\it e.g.}, Refs.\,\cite{Ril19a, Spi22a, Spi23a}. Known decay branching for excited states of \nuc{72,74}{Se} \cite{ENSDF, Sin06a, Abr10a, McC11a, Muk22a} was explicitly taken into account by using the \textsc{geant4} photo-evaporation database format \cite{geant4} implemented in \textsc{ucgretina}. This procedure also allowed for the correction of the $\gamma$-ray yields for observed feeders (see also Refs.\,\cite{Spi22a, Spi23a}). Partial cross sections were calculated from the experimental $\gamma$-ray yields by normalizing these to the number of incoming beam particles and the number of target nuclei. For the inclusive cross sections, the number of outgoing \nuc{72,74}{Se} reaction residues measured with the S800 spectrograph were used instead of the $\gamma$-ray yields. As first described in Ref.\,\cite{Ril19a}, the LH$_2$ target thickness was determined via a comparison of the measured kinetic-energy distribution of the reacted outgoing beam to a detailed \textsc{geant4} simulation performed with \textsc{ucgretina} \cite{Ril21a}. The simulation also uses the independently measured kinetic-energy distribution of the incoming beam through the empty target cell as input. An areal target density of 69(3) mg/cm$^2$ was determined\,\cite{Spi22a, Spi23a} resulting in mid-target beam energies of $\sim 89$\,MeV/u for \nuc{75}{Br} and $\sim 95$\,MeV/u for \nuc{73}{Br}, respectively. No acceptance losses in the momentum distributions were observed. Wimmer {\it et al.} stated that the average mid-target energy for their proton-knockout experiment on a \nuc{9}{Be} nuclear target was $\sim 140$\,MeV/u\,\cite{Wim18a}.

\renewcommand*{\arraystretch}{1.3}
\begin{longtable}[t]{cccccccc}
\caption{\label{table_01} Experimental data for \nuc{72}{Se}. Given are the adopted excitation energy $E_{x}$ of the observed (populated) states, their spin-parity assignment $J^{\pi}_i$, the energies $E_{\gamma}$ for $\gamma$-ray transitions observed from these states, the excitation energy $E_{f}$ and spin-parity assignment $J^{\pi}_f$ of states these $\gamma$-ray transitions lead to, the previously reported as well as the $\gamma$-decay intensity used for the \textsc{ucgretina} simulation, and the partial cross section $\sigma_{\mathrm{part.}}$ determined for proton removal. The latter are corrected for known, observed feeders. Information from previous experiments and adopted data were taken from Refs.\,\cite{ENSDF, Sin06a, Abr10a, McC11a, Muk22a}. If not noted otherwise, the adopted spin-parity assignments and $\gamma$-decay branching ratios\,\cite{ENSDF} are given in the table. The summed partial cross section for resolved excited states of \nuc{72}{Se} is given at the bottom of the table.}
\vspace{2mm}
\\
\hline
\hline
\multicolumn{1}{c}{$E_{x}$} & \multicolumn{1}{c}{$J^{\pi}_i$} & \multicolumn{1}{c}{$E_{\gamma}$} & \multicolumn{1}{c}{$E_{f}$} & \multicolumn{1}{c}{$J^{\pi}_f$} & \multicolumn{2}{c}{$I_{\gamma}$\,[$\%$]} & \multicolumn{1}{c}{$\sigma_{\mathrm{part.}}$} \\
\cline{6-7}
\multicolumn{1}{c}{[keV]} & & \multicolumn{1}{c}{[keV]} & \multicolumn{1}{c}{[keV]} & & \multicolumn{1}{c}{Ref.\,\cite{Abr10a}} & \multicolumn{1}{c}{This work} & \multicolumn{1}{c}{[mb]} \\
\hline
\hline
\endfirsthead
\multicolumn{8}{c}{Table \ref{table_01}: ({\it Continued.)}}
\\
\hline
\hline
\multicolumn{1}{c}{$E_{x}$} & \multicolumn{1}{c}{$J^{\pi}_i$} & \multicolumn{1}{c}{$E_{\gamma}$} & \multicolumn{1}{c}{$E_{f}$} & \multicolumn{1}{c}{$J^{\pi}_f$} & \multicolumn{2}{c}{$I_{\gamma}$\,[$\%$]} & \multicolumn{1}{c}{$\sigma_{\mathrm{part.}}$} \\
\cline{6-7}
\multicolumn{1}{c}{[keV]} & & \multicolumn{1}{c}{[keV]} & \multicolumn{1}{c}{[keV]} & & \multicolumn{1}{c}{Ref.\,\cite{Abr10a}} & \multicolumn{1}{c}{This work} & \multicolumn{1}{c}{[mb]} \\
\hline
\hline
\endhead
\hline
\endfoot
\hline
\hline
\multicolumn{8}{c}{${}^{a}$ Ref.\,\cite{Muk22a} lists $J^{\pi} = 4^+$ and established rotational band.} \\
\multicolumn{8}{c}{${}^{b}$ In agreement with Refs.\,\cite{McC11a, Muk22a}.} \\
\multicolumn{8}{c}{${}^{c}$ Observed in Ref.\,\cite{McC11a}.} \\
\endlastfoot

\multicolumn{8}{c}{\nuc{72}{Se}; $\sigma_{\mathrm{incl.}} = 68(4)$\,mb} \\
862  & $2^+_1$ & 862  & 0    & $0^+_1$ & 100       & 100 & 8.9(8) \\
1317 & $2^+_2$ & 380  & 937  & $0^+_2$ &  16.6(14) & 8   & 6.0(7)  \\
     &         & 455  & 862  & $2^+_1$ &  36(3)    & 40 &        \\
     &         & 1317 & 0    & $0^+_1$ &  47(4)    & 52 &        \\
1637 & $4^+_1$ & 775  & 862  & $2^+_1$ & 100       & 100 & 7.2(8)  \\
1876 & $(2,4)$ & 559  & 1317 & $2^+_2$ & 79(13)    & 79  & 1.3(3)  \\
     &         & 1014 & 862  & $2^+_1$ & 21(12)    & 21  &        \\
1999 & $2^+$   & 1062 & 937  & $0^+_2$ & 44(7)     & 52  & 0.5(4)  \\
     &         & 1137 & 862  & $2^+_1$ & 56(7)     & 48  &        \\
2294 & $(2)^a$  & 977  & 1317 & $2^+_2$ & 53(3)     & 46  & 2.2(4)  \\
     &         & 1432 & 862  & $2^+_1$ & 47(5)     & 54  &        \\
2406 & $3^-_1$ & 1089 & 1317 & $2^+_2$ & 100       & 100 & 1.7(3)  \\
2434 & $3^-_2$ & 1117 & 1317 & $2^+_2$ & 16(2)     & 9$^b$   & 2.7(3)  \\
     &         & 1572 & 862  & $2^+_1$ & 63(5)     & 91$^b$ &    \\
     &         & 2434 & 0    & $0^+_1$ & 21(5)     & 0$^b$  &    \\
2467 & $6^+_1$ & 830  & 1637 & $4^+_1$ & 100       & 100 & 5.8(5)  \\
2586 & $(3)$   & 710  & 1876 & $(2,4)$ & 27(8)     & 24  & 0.8(5)  \\
     &         & 1269 & 1317 & $2^+_2$ & 14(7)     & 8   &    \\
     &         & 1724 & 862  & $2^+_1$ & 58(11)    & 68  &    \\
3094$^c$ &     & 800  & 2294 & $(2)$   & 31(6)     & 31  & 4.6(6)  \\
     &         & 1095 & 1999 & $2^+$   & 46(10)    & 46  &    \\
     &         & 1457 & 1637 & $4^+_1$ & 23(5)     & 23  &    \\
3173 & $5^-_1$ & 740  & 2434 & $3^-_2$ & 14(3)     & 18  & 6.5(5)  \\
     &         & 879  & 2294 & $(2)$   & $\leq 10$ & 15  &    \\
     &         & 1536 & 1637 & $4^+_1$ & 76(11)    & 67  &    \\
3350 & $5^-$   & 916  & 2434 & $3^-_2$ & 10(2)     & 11  & 1.0(4)  \\
     &         & 1713 & 1637 & $4^+_1$ & 90(10)    & 89  &    \\
3425 & $8^+_1$ & 958  & 2467 & $6^+_1$ & 100       & 100 & 1.8(2)  \\
\hline
\multicolumn{8}{r}{$\sum \sigma_{\mathrm{part.}} = 51(7)$\,mb} \\
\end{longtable}

\section{Results and Discussion}
 
\begin{figure}[t]
\centering
\includegraphics[width=1\linewidth]{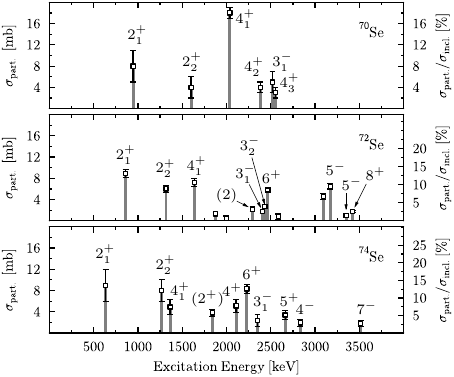}
\caption{\label{fig:crosssections}{Partial cross sections for excited states of \nuc{70,72,74}{Se} populated in proton knockout from \nuc{71,73,75}{Br}. The \nuc{70}{Se} data are taken from Ref.\,\cite{Wim18a}. Spin-parity assignments are specified if known from previous experiments\,\cite{ENSDF, Sin06a, Abr10a, McC11a, Muk22a}. The proton-separation energies, $S_p$, are at 6110(30), 7264(5), and 8549(4)\,keV for \nuc{70,72,74}{Se}\,\cite{ENSDF}, respectively. The neutron-separation energies, $S_n$, are significantly higher in all three isotopes. For illustrative purposes, the ratio between the partial cross sections and inclusive cross section is shown on the second $y$ axis.}} 
 \end{figure}

The experimental results are shown in Tables \ref{table_01} and \ref{table_02} as well as in Fig.\,\ref{fig:crosssections}. Inclusive cross sections, $\sigma_{\mathrm{incl.}}$, of 68(4) and 66(4)\,mb were determined for \nuc{72}{Se} and \nuc{74}{Se}, respectively. Stated uncertainties include statistical uncertainties, the stability of the secondary beam composition, uncertainties coming from the choice of software gates and the target thickness. The two inclusive cross sections on the proton target agree within uncertainties. This might suggest that the same single-particle orbitals contribute to the proton-removal reaction. To further investigate this hypothesis, we take a closer look at the population of excited states of the neutron-deficient Se isotopes in proton removal. 

\begingroup
\squeezetable
\renewcommand*{\arraystretch}{1.2}
\begin{table}[!t]
\caption{\label{table_02} Same as Table\,\ref{table_01} but for \nuc{74}{Se}. The 2537-keV and 2900-keV transitions could not be placed in the level scheme (see Fig.\,\ref{fig:spectra}).}
\begin{ruledtabular}
\begin{tabular}{cccccccc}

\multicolumn{1}{c}{$E_{x}$} & \multicolumn{1}{c}{$J^{\pi}_i$} & \multicolumn{1}{c}{$E_{\gamma}$} & \multicolumn{1}{c}{$E_{f}$} & \multicolumn{1}{c}{$J^{\pi}_f$} & \multicolumn{2}{c}{$I_{\gamma}$\,[$\%$]} & \multicolumn{1}{c}{$\sigma_{\mathrm{part.}}$} \\
\cline{6-7}
\multicolumn{1}{c}{[keV]} & & \multicolumn{1}{c}{[keV]} & \multicolumn{1}{c}{[keV]} & & \multicolumn{1}{c}{Ref.\,\cite{Sin06a}} & \multicolumn{1}{c}{This work} & \multicolumn{1}{c}{[mb]} \\
\hline

\multicolumn{8}{c}{\nuc{74}{Se}; $\sigma_{\mathrm{incl.}} = 66(4)$\,mb} \\
635  & $2^+_1$ & 635  & 0    & $0^+_1$ & 100       & 100 & 9(3) \\
1269 & $2^+_2$ & 634  & 635  & $2^+_1$ &  66(7)    & 66  & 8(2)  \\
     &         & 1269 & 0    & $0^+_1$ &  34(3)    & 34  &        \\
1363 & $4^+_1$ & 728  & 635  & $2^+_1$ & 100       & 100 & 4.9(14) \\
1839 & $(2^+)$ & 985  & 854  & $0^+_2$ &  82(12)   & 82  & 3.8(7)  \\
     &         & 1204 & 635  & $2^+_1$ &  18(9)    & 18  &        \\
2108 & $4^+$   & 745  & 1363 & $4^+_1$ &  24(3)    & 27  & 5.1(12)  \\
     &         & 839  & 1269 & $2^+_2$ &  61(7)    & 56  &        \\
     &         & 1473 & 635  & $2^+_1$ &  15(2)    & 17  &        \\
2231 & $6^+_1$ & 868  & 1363 & $4^+_1$ & 100       & 100 & 8.3(9) \\
2350 & $3^-_1$ & 987  & 1363 & $4^+_1$ &  23(5)    & 20  & 2.4(11) \\
     &         & 1081 & 1269 & $2^+_2$ &  40(8)    & 37  &        \\
     &         & 1715 & 635  & $2^+_1$ &  37(6)    & 43  &        \\
2662 & $5^+$   & 778  & 1884 & $3^+_1$ &  68(12)   & 68  & 3.4(8) \\
     &         & 1299 & 1363 & $4^+_1$ &  32(12)   & 32  &        \\
2832 & $4^-$   & 1469 & 1363 & $4^+_1$ &  100      & 100 & 1.9(7) \\
3516 & $7^-$   & 529  & 2987 & $6^+$   &  $\leq 4$ & 0   & 1.8(6) \\
     &         & 673  & 2843 & $5^-$   &  89(10)   & 93  &        \\
     &         & 1285 & 2231 & $6^+_1$ &  7.1(10)  & 7   &        \\
     \hline
\multicolumn{8}{r}{$\sum \sigma_{\mathrm{part.}} = 48(13)$\,mb\footnote{Not including the unplaced 2537-keV and 2900-keV transitions. For these, $\sigma_{\mathrm{part.}}$ would be 2.0(7) and 1.1(6)\,mb, respectively.}} \\

\end{tabular}
\end{ruledtabular}
\end{table}
\endgroup

The partial cross sections, $\sigma_{\mathrm{part.}}$, determined for resolved excited states of \nuc{70-74}{Se} are compared in Fig.\,\ref{fig:crosssections}. They are remarkably similar in magnitude. The large partial cross section to the $4^+_1$ state of \nuc{70}{Se} is noteworthy. It was already mentioned in Ref.\,\cite{Wim18a}. With our new data, we see that this cross section increases gradually from 4.9(14)\,mb [7(2)\,$\%$] in \nuc{74}{Se}, over 7.2(8)\,mb [10.5(13)\,$\%$] in \nuc{72}{Se}, to 18(1)\,mb [18(2)\,$\%$] in \nuc{70}{Se}. Here, we state the ratio of the partial cross section relative to the inclusive cross section in square brackets. This ratio is less sensitive to differences in the reaction kinematics between proton removal from nuclear and nucleon targets than the absolute partial cross section values. In Ref.\,\cite{Wim18a}, Wimmer {\it et al.} mentioned that they performed cross-section calculations using shell-model input from the JUN45 interaction. They found that the cross section to both the \nuc{70}{Se} ground state and $4^+_1$ state were largest if the ground state of \nuc{71}{Br} was assumed to be a $5/2^-$ state. They did not provide any values for the expected magnitude of these theoretical cross sections. Nonetheless, as the ground states of \nuc{73,75}{Br} have firm spin-parity assignments of $J^{\pi} = 3/2^-$ and $1/2^-$, respectively, this hypothesis seems plausible. There is a caveat though. In our work, we observe the population of the $6^+_1$ state, which is the strongest feeder of the $4^+_1$ state. Between the strong 945-keV, $2^+_1 \rightarrow 0^+_1$ and 1094-keV, $4^+_1 \rightarrow 2^+_1$ transitions, it is possible that the 964-keV, $6^+_1 \rightarrow 4^+_1$ transition of \nuc{70}{Se} could not be resolved in the \nuc{9}{Be}(\nuc{71}{Br},X)\nuc{70}{Se} experiment (see Fig.\,5 in Ref. \cite{Wim18a}). The partial cross section to the $4^+_1$ state in \nuc{70}{Se} might, therefore, be significantly smaller. This possibility reinforces the importance of using high-resolution $\gamma$-ray arrays for experiments with deformed nuclei. We will discuss the population of the $6^+_1$ state shortly. For completeness, we provide upper limits for the experimentally determined ground state partial cross sections, which can be obtained by subtracting the sum of the partial cross sections to resolved excited states from the inclusive cross section: 18(13)\,mb [26(19)\,$\%$] in \nuc{74}{Se} and 17(8)\,mb [26(12)\,$\%$] in \nuc{72}{Se}. For \nuc{70}{Se}, Wimmer {\it et al.} reported 51(18)\,mb [51(19)\,$\%$]. It appears that the cross section to the ground state might also be the largest for \nuc{70}{Se} and, thus, would support the argument made by Wimmer {\it et al.} \cite{Wim18a}. We chose to not add these values to Tables \ref{table_01} and \ref{table_02} though as they have to be considered upper limits. Note that the 2537-keV and 2900-keV transitions in \nuc{74}{Se} could not be placed. If they led directly to the ground state, then the stated upper limit would already be lower. When inspecting Fig.\,\ref{fig:crosssections}, it is also quite obvious that several additional states are observed in both \nuc{72}{Se} and \nuc{74}{Se}, which could either not be resolved in the \nuc{70}{Se} experiment or which were not populated in the proton-removal reaction on the \nuc{9}{Be} target. If they could not be resolved and some decay to the ground state, then the partial cross section to the ground state of \nuc{70}{Se} would also be smaller.

\begin{figure}[t]
\centering
\includegraphics[width=1\linewidth]{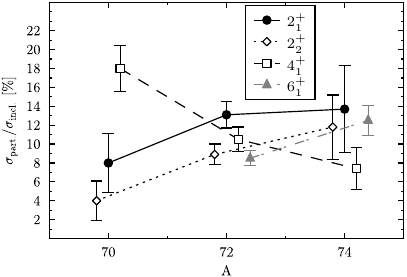}
\caption{\label{fig:crosssections_rel}{Partial cross sections relative to the inclusive cross section for the excited $2^+_1$, $2^+_2$, $4^+_1$, and $6^+_1$ states of \nuc{70,72,74}{Se} populated in proton knockout from \nuc{71,73,75}{Br}. The \nuc{70}{Se} data are taken from Ref.\,\cite{Wim18a}. See text for further discussion.}} 
 \end{figure}

In Fig.\,\ref{fig:crosssections_rel}, we compare the relative cross sections for the $2^+_1$, $2^+_2$, $4^+_1$, and $6^+_1$ states. As mentioned, these relative cross sections are less sensitive to reaction-kinematics differences between proton removal from nuclear and nucleon targets than the absolute partial cross section values. Given the present uncertainties, it is quite clear that no specific trend can be claimed; even though it is tempting to see a decrease of the $2^+_1$ and $2^+_2$ population with decreasing mass $A$, and to claim an increase of the relative cross section for the $4^+_1$ state with decreasing mass $A$. The relative population of the $6^+_1$ might also decrease. We will pick this up when discussing the population of this state below. Without guidance by theory, it is not clear whether these possible trends for the partial cross section ratios can be reconciled with a pronounced prolate to oblate shape change between \nuc{72}{Se} and \nuc{70}{Se} \cite{Lju08a, Hen18a}, or whether they can be simply attributed to a changing ground state spin between the Br isotopes. Theoretical calculations of the partial cross sections using the eikonal reaction theory with shell-model input, possibly implementing the full $fp$ shell, the $1g_{9/2}$ orbital as well as the more deeply bound $sd$ orbitals, could be useful. As beyond-mean-field approaches based on nuclear density functional theory have also provided important input to understand the structure of these complex nuclei in recent years, it might be instructive to calculate spectroscopic factors needed as input for the reaction calculations from these theoretical approaches, too. We acknowledge that such calculations are challenging at the moment. 

 \begin{figure}[t]
\centering
\includegraphics[width=1\linewidth]{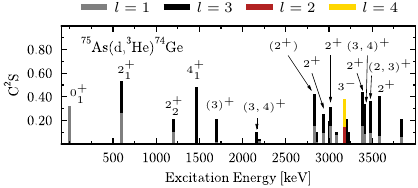}
\caption{\label{fig:74ge}{Spectroscopic factors for excited states of \nuc{74}{Ge} populated through $l = 1, 2, 3, 4$ proton removal in the \nuc{75}{As}$(d,\nuc{3}{He})$\nuc{74}{Ge} reaction at $E_d = 26$\,MeV \cite{Rot77a}. For each state, the different contributions have been summed up. The length of the colored bars corresponds to the individual contribution of each orbital, respectively. The larger contribution is shown on top of the smaller one. The summed spectroscopic strength for $l=1$ and $l=3$ is 5.3. The ground state spin-parity assignment of \nuc{75}{As} ($Z = 33, N = 42$) is $J^{\pi} = 3/2^-$, i.e., the same as for \nuc{75}{Br} ($Z= 35, N = 40$).}} 
 \end{figure}

\begin{figure}[t]
\centering
\includegraphics[width=1\linewidth]{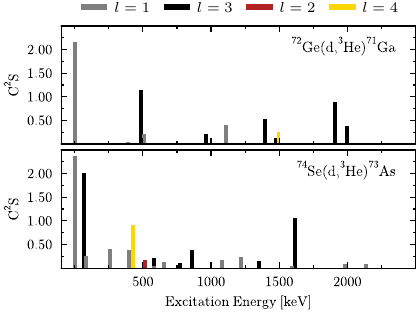}
\caption{\label{fig:n40}{Spectroscopic factors for excited states of the $N=40$ isotones \nuc{71}{Ga} (top) and \nuc{73}{As} (bottom) populated through $l = 1, 2, 3, 4$ proton removal in the $(d,\nuc{3}{He})$ reaction at $E_d = 26$\,MeV from targets of \nuc{72}{Ge} and \nuc{74}{Se}, respectively \cite{Rot78a, Rot83a}. The $l=1, 3$ and 4 strengths shift significantly down in \nuc{73}{As}.}} 
 \end{figure}

However, available data for excited states of \nuc{74}{Ge} populated through the \nuc{75}{As}$(d,\nuc{3}{He})$\nuc{74}{Ge} proton-removal reaction in normal kinematics \cite{Rot77a} suggest indeed that all of the orbitals mentioned above contribute. We compiled this information in Fig.\,\ref{fig:74ge}. As can be seen, several states were populated including the 3175-keV, $J^{\pi} = 3^-$ state. The latter can only be populated through proton removal from orbitals with angular momentum $l=2$ or $l=4$. Interestingly, at least three other firmly assigned $3^-$ states are known below the 3175-keV state \cite{ENSDF}. In the Se isotopes, the $3^-_1$ states were populated in one-proton removal. In addition, the $3^-_2$ state was populated in \nuc{72}{Se} (see Fig.\,\ref{fig:crosssections}). It seems that the configurations leading to the population of these states can be found at lower energies in the Se isotopes than in \nuc{74}{Ge}. At this moment, it is not clear whether this possible shift in energy might also contribute to the significant increase of octupole collectivity observed around mass $A = 72$ \cite{Spi22a}. The observation appears to be in line with proton-removal data from even-$A$ to odd-$A$ nuclei in this mass range though, where the $l=2$ and $l=4$ strengths were also observed at lower excitation energies for reactions on Se targets than on Ge targets \cite{Rot78a, Rot83a}. In addition, more of the $l=1$ and $l=3$ strengths were observed to be concentrated in lower-lying states in the \nuc{A}{Se}$(d,\nuc{3}{He})$\nuc{A-1}{As} than in the \nuc{A}{Ge}$(d,\nuc{3}{He})$\nuc{A-1}{Ga} reactions \cite{Rot78a, Rot83a} (see Fig.\,\ref{fig:n40} for the $N=40$ isotones \nuc{71}{Ga} and \nuc{73}{As}). A general shift of the spectroscopic strengths to lower energies might, thus, explain why we do not observe higher-lying, positive-parity states in the Se isotopes. In this context, we want to emphasize that, even though we cannot fully exclude feeding contributions from unresolved excited states at higher energies, we would have been able to detect the $\gamma$ decay of states as strongly populated as those above 2.5\,MeV in \nuc{74}{Ge} (see Fig.\,\ref{fig:74ge}) with GRETINA. In \nuc{74}{Ge}, all of these states decay with $\gamma$-ray energies of less than 1.5\,MeV \cite{ENSDF}. The residuals between the simulated and experimental spectra up to 3\,MeV suggest that we do not miss significant, additional strength up to 3\,MeV (see Fig.\,\ref{fig:spectra}). It needs to be acknowledged, however, that the unplaced 2537-keV and 2900-keV transitions are barely resolved in our present experiment. The determined yields would correspond to partial cross sections of 2.0(7) and 1.1(6)\,mb (see footnote of table\,\ref{table_02}), respectively. They certainly appear to be among the smaller partial cross sections we determined (see Fig.\,\ref{fig:crosssections}). We can, thus, justifiably assume that we should have been able to resolve states with larger cross sections than these.
 
Additionally, it must be mentioned that, as Wimmer {\it et al.} \cite{Wim18a}, we cannot entirely exclude that some component of the incoming secondary beam was in an isomeric state. However, in all three cases, the half-lives of the known isomers are less than 40\,ns\,\cite{ENSDF}. Consequently, a significant fraction of this possible isomeric component should have decayed when reaching the reaction target. Nonetheless, both \nuc{71}{Br} and \nuc{73}{Br} have very low lying excited states at 9.9 and 26.9\,keV, respectively, with unknown lifetimes, which we need to consider. If we assume that the $B(E2;5/2^- \rightarrow 1/2^-) \approx 3.8$\,W.u. as in \nuc{71}{Se}\,\cite{ENSDF}, then a lifetime on the order of 800\,ns would be expected for the 27-keV state in \nuc{73}{Br}. An isomeric state with such a lifetime could contribute to the incoming secondary beam. Thus, we want to point out again that the inclusive cross sections for \nuc{72}{Se} and \nuc{74}{Se} are identical within uncertainties. Furthermore, the general partial cross section pattern is very similar both in cross section fragmentation as well as in absolute magnitude (see Fig.\,\ref{fig:crosssections}). Based on these observations and on the fact that \nuc{75}{Br} does not have such a low lying isomeric state \cite{ENSDF}, we claim that the influence of a possible isomeric component of the beam is likely minor. It is, therefore, surprising that the $6^+$ state was populated with an appreciable cross section in both proton removal from \nuc{73}{Br} and \nuc{75}{Br}, for which the ground-state spins are known to be $J < 5/2$. Direct proton knockout from the $1f_{7/2}$ orbital could explain the population of the $6^+$ state in \nuc{72,74}{Se} if the ground-state spin of the corresponding Br isotope was $J=5/2$ as seems to be the case in \nuc{71}{Br}. We mentioned earlier that the $6^+_1 \rightarrow 4^+_1$ transition of \nuc{70}{Se} could possibly not be resolved in Ref.\,\cite{Wim18a}. In our case, we even observe the weak population of an excited $7^-$  state in \nuc{74}{Se} and of the probably first $8^+$ state in \nuc{72}{Se}. The population of these comparably high-$J$ states might indicate that multi-step processes contribute. The possibility that such processes could contribute to nucleon-removal reactions at intermediate energies was also discussed in, {\it e.g.}, Refs. \cite{Poh23a}. Since excited states belonging to rotational bands in both the Br and Se isotopes are connected through collective $E2$ matrix elements\,\cite{ENSDF}, states could be populated to some extent through two-step processes. The degree of quadrupole deformation, $\beta_2$, decreases by 74(3)\,$\%$ from \nuc{74}{Se} to \nuc{72}{Se} \cite{Pri14a}. Thus, one might expect that two-step processes are weaker in the proton removal from \nuc{73}{Br} to \nuc{72}{Se}. Coincidently, the partial cross section ratio $\sigma(6^+_1)_{\nuc{72}{Se}}/\sigma(6^+_1)_{\nuc{74}{Se}}$ is 68(11)\,$\%$. At the moment, such indirect contributions are out of the scope of the eikonal reaction theory. We can, thus, not conclude whether there is a connection even though the ratios are the same within uncertainties. Further reaction theory developments along the lines of Ref. \cite{Poh23a} would be needed. For completeness, we note that indirect processes contributed significantly to the population of excited states of the strongly deformed nuclei \nuc{168}{Er} and \nuc{240}{Pu} in the $(p,t)$ reaction at 25\,MeV \cite{Spi18a, Buc23a}. Contributions of multi-step processes to proton knockout on nuclear targets were also discussed in, {\it e.g.}, Refs.\,\cite{Spi19a,Jon20a,Aum21a}.

\section{Summary}

We performed proton-removal experiments on \nuc{73,75}{Br} secondary beams populating excited states of \nuc{72,74}{Se}. The inclusive cross sections for both Se isotopes agree within uncertainties. This could suggest that the same single-particle orbitals contribute to the proton-removal reaction. However, without detailed theoretical structure and reaction calculations this statement cannot be backed up. Using the high-resolution GRETINA $\gamma$-ray tracking array, several excited states could be identified. We discussed the fragmentation of the partial cross section among these excited states. But without guidance by theory, it is not clear whether possible trends of the cross-section ratios originate from a pronounced prolate to oblate shape change between \nuc{72}{Se} and \nuc{70}{Se}, or whether they can be simply attributed to a changing ground state spin between the Br isotopes. We, thus, call for combined and detailed structure and reaction calculations. Such calculations are challenging at the moment and, possibly, further theoretical developments are needed. A comparison to proton-removal data available from $(d,\nuc{3}{He})$ reactions on stable isotopes suggests, however, that a significant fraction of the $l= 1, 2, 3$ and 4 spectroscopic strengths might be found at lower energies in the neutron-deficient Se isotopes than the Ge isotopes. The situation might be further complicated since we presented some evidence that, in these collective nuclei, multi-step processes contribute to the population of excited states through proton removal on nucleon targets at intermediate energies. At the moment, such multi-step contributions are out of the scope of the eikonal reaction theory. Further reaction theory developments along the lines of Ref. \cite{Poh23a} seem instructive as many of the heavier nuclei, which will become accessible for experiments at the Facility for Rare Isotope Beams (FRIB), will probably be deformed. It is likely that nucleon-removal experiments will be performed with both nucleon and nuclear targets. Extending the studies of Refs.\,\cite{Tos21b, Wan23a} to heavier systems might, thus, be instructive.

%The inclusive cross section ratio with respect to the inclusive cross section measured on the \nuc{9}{Be} target for \nuc{70}{Se} was found to be 67(9)\,$\%$. As mentioned in the text, the secondary-beam energies were different though. 

%Calculations performed with currently available shell-model interactions in the $fp$ and $fpg$ model spaces can, however, not support this hypothesis as they fail to reproduce the observed fragmentation of the spectroscopic strength and the magnitude of the experimentally determined partial cross sections to excited states in the neutron-deficient Se isotopes. Even though contributions from two-step processes seem important in these complex, collective nuclei, further theoretical developments are needed to provide more robust shell-model spectroscopic factors for the eikonal reaction theory calculations to be able to reliably study such effects.

\begin{acknowledgments}
This work was supported by the National Science Foundation (NSF) under Grant No. PHY-2012522 (WoU-MMA: Studies of Nuclear Structure and Nuclear Astrophysics), Grant No. PHY-1565546 (NSCL), Grant No. PHY-2209429 (Windows on the Universe: Nuclear Astrophysics at FRIB), and by the Department of Energy, Office of Science, Office of Nuclear Physics, Grant No. DE-SC0023633 (MSU). GRETINA was funded by the Department of Energy, Office of Science. The operation of the array at NSCL was supported by the DOE under Grant No. DE-SC0019034. M.S. acknowledges support through the FRIB Visiting Scholar Program for Experimental Science 2020. M.S. is also grateful to B.A. Brown, J.A. Tostevin, and A. Volya for continuing discussions about nuclei, structure, and reactions in the Ge-Zr mass region as well as for providing additional insights through schematic shell-model plus eikonal reaction theory calculations. 
\end{acknowledgments}

%M.S. is grateful to B.A. Brown and A. Volya for continuing discussions about nuclei and structure in the Ge-Zr mass region and for providing additional insight through schematic shell-model calculations.
\bibliography{72_74Se_pKO}

\end{document}